\documentclass[12pt]{article}
\usepackage{epsfig}       

\newcommand{\ba}{\begin{array}}
\newcommand{\ea}{\end{array}}

\newcommand{\pl}{P_L}
\newcommand{\pr}{P_R}

\newcommand{\limit}[2]{{\mbox{
                        $\longrightarrow$\hskip 0cm\kern-.6cm
                        \vskip -.25cm
                        ${\scriptscriptstyle #1\rightarrow #2}$}
                     }}

\newcommand{\tb}{\ensuremath{\tan\beta}}
\newcommand{\tbs}{\ensuremath{\tan^2\!\beta}}
\newcommand{\ctb}{\ensuremath{\cot\beta}}
\newcommand{\ctbs}{\ensuremath{\cot^2\!\beta}}
\newcommand{\sbt}{\ensuremath{\sin\beta}}

\newcommand{\cbt}{\ensuremath{\cos\beta}}

\newcommand{\sa}{\ensuremath{\sin\alpha}}

\newcommand{\ca}{\ensuremath{\cos\alpha}}

\newcommand{\sbma}{\ensuremath{\sin (\beta-\alpha)}}
\newcommand{\cbma}{\ensuremath{\cos (\beta-\alpha)}}

\newcommand{\mb}{\ensuremath{m_b}}
\newcommand{\mbs}{\ensuremath{m_b^2}}
\newcommand{\mt}{\ensuremath{m_t}}
\newcommand{\mts}{\ensuremath{m_t^2}}

\newcommand{\mw}{\ensuremath{M_W}}
\newcommand{\mws}{\ensuremath{M^2_W}}
\newcommand{\mz}{\ensuremath{M_Z}}

\newcommand{\mh}{\ensuremath{M_{h^0}}}
\newcommand{\mH}{\ensuremath{M_{H^0}}}
\newcommand{\mHp}{\ensuremath{M_{H^\pm}}}
\newcommand{\mA}{\ensuremath{M_{A^0}}}

\newcommand{\mHs}{\ensuremath{M^2_{H^0}}}
\newcommand{\mHps}{\ensuremath{M^2_{H^\pm}}}
\newcommand{\mAs}{\ensuremath{M^2_{A^0}}}

\newcommand{\mg}{\ensuremath{m_{\tilde{g}}}}

\newcommand{\msbo}{\ensuremath{m_{\tilde{b}_1}}}

\newcommand{\msto}{\ensuremath{m_{\tilde{t}_1}}}

\newcommand{\Bz}{\ensuremath{B_0}}
\newcommand{\Bo}{\ensuremath{B_1}}

\newcommand{\Cz}{\ensuremath{C_0}}
\newcommand{\Coo}{\ensuremath{C_{11}}}
\newcommand{\Cot}{\ensuremath{C_{12}}}
\newcommand{\Czt}{\ensuremath{\tilde{C}_0}}

\newcommand{\tbsH}{\ensuremath{\frac{a_{j}}{\cot\!\beta}}}
\newcommand{\ctbsH}{\ensuremath{\frac{\cot\!\beta}{a_{j}}}}
\newcommand{\sab}{\ensuremath{\sin\! (\alpha-\beta)}}
\newcommand{\cab}{\ensuremath{\cos\! (\alpha-\beta)}}
\newcommand{\sasH}{\ensuremath{r_{j}^{2}}}
\newcommand{\casH}{\ensuremath{R_{j}^{2}}}

\renewcommand{\mh}{M_{H^{+}}}

\newcommand{\maz}{M_{A^0}}
\newcommand{\mHz}{M_{H^0}}
\newcommand{\mhz}{M_{h^0}}
\newcommand{\mhzs}{M_{h^0}^2}

\newcommand{\beq}{\begin{equation}}
\newcommand{\eeq}{\end{equation}}
\newcommand{\beqn}{\begin{eqnarray}}
\newcommand{\eeqn}{\end{eqnarray}}

\newcommand{\stackm}{\stackrel{\scriptstyle <}{{ }_{\sim}}}

\newcommand{\figtbthdm}{
\begin{figure}
\begin{center}
\epsfig{file=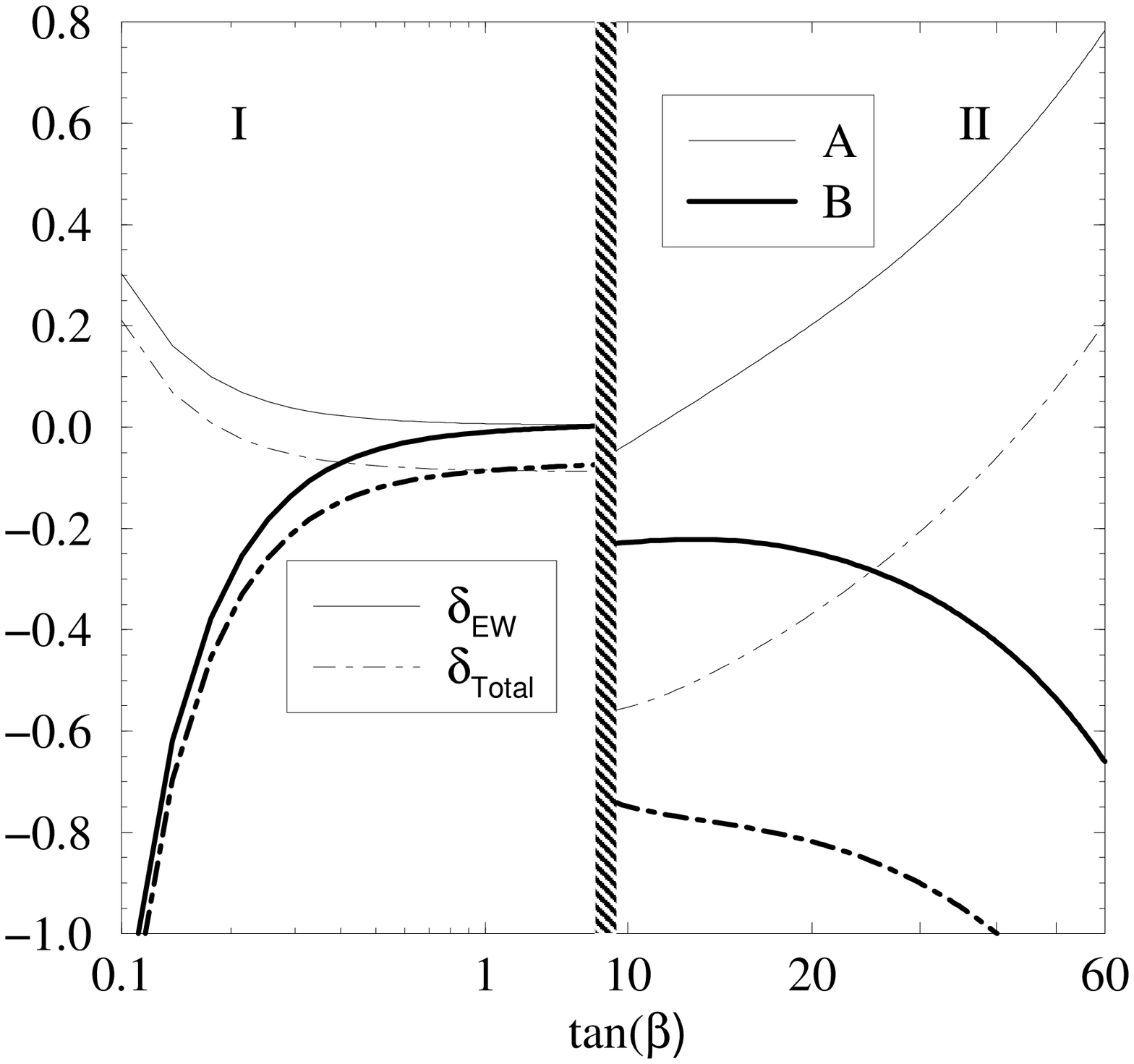,width=7cm}
\end{center}
\caption{ The correction $\delta$, eq.\,(\ref{eq:defdelta}),
  to the decay width
  $\Gamma(t \rightarrow H^+ b)$ as a function of $\tan\beta$,
  for Type I $2$HDM's (left hand side of the figure)
  and two sets of inputs $\{(\mh,\mHz,\mhz,\maz);\, \tan\alpha\}$, 
  namely
  set A: $\{(70,175,100,50)\,GeV;\,3\}$ and
  set B: $\{(120,200,80,250)\,GeV; 1\}$.
  Similarly for Type II models (right hand side of the figure)
  and for two different sets of inputs,
  set A: $\{(120,300,50,225)\,GeV;\,1\}$ and set
  B: $\{(120,300,80,225)\,GeV;\,-3\}$.
  Shown are the electroweak contribution $\delta_{\rm EW}$ and the
  total correction $\delta_{\rm Total}=\delta_{\rm EW}+
  \delta_{\rm QCD}$.}\label{fig:deltatb2hdm}
\end{figure}
}

\newcommand{\figtbmssm}{\begin{figure}
\begin{center}
\begin{tabular}{cc}
\epsfig{file=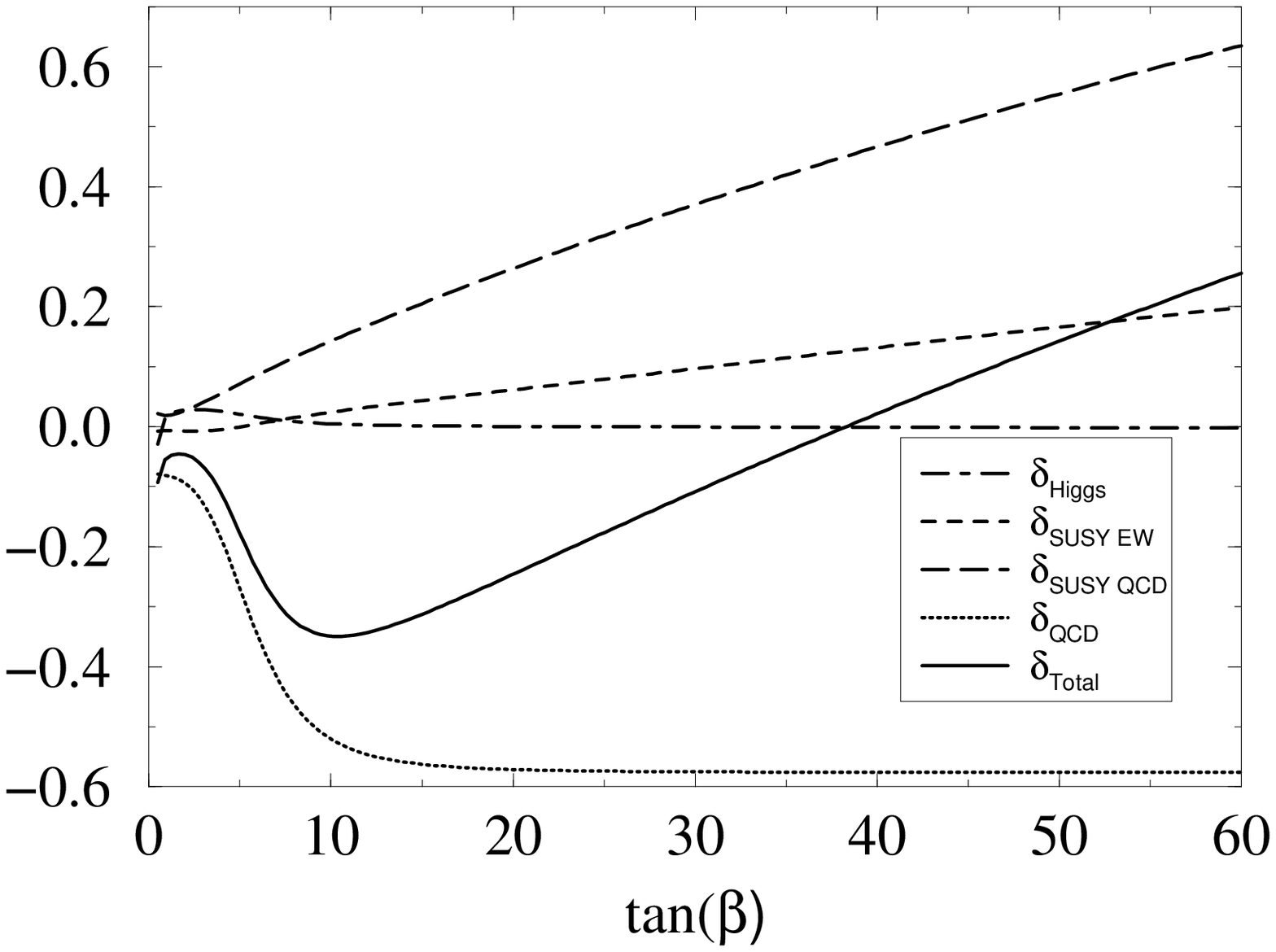,width=7cm} & 
\epsfig{file=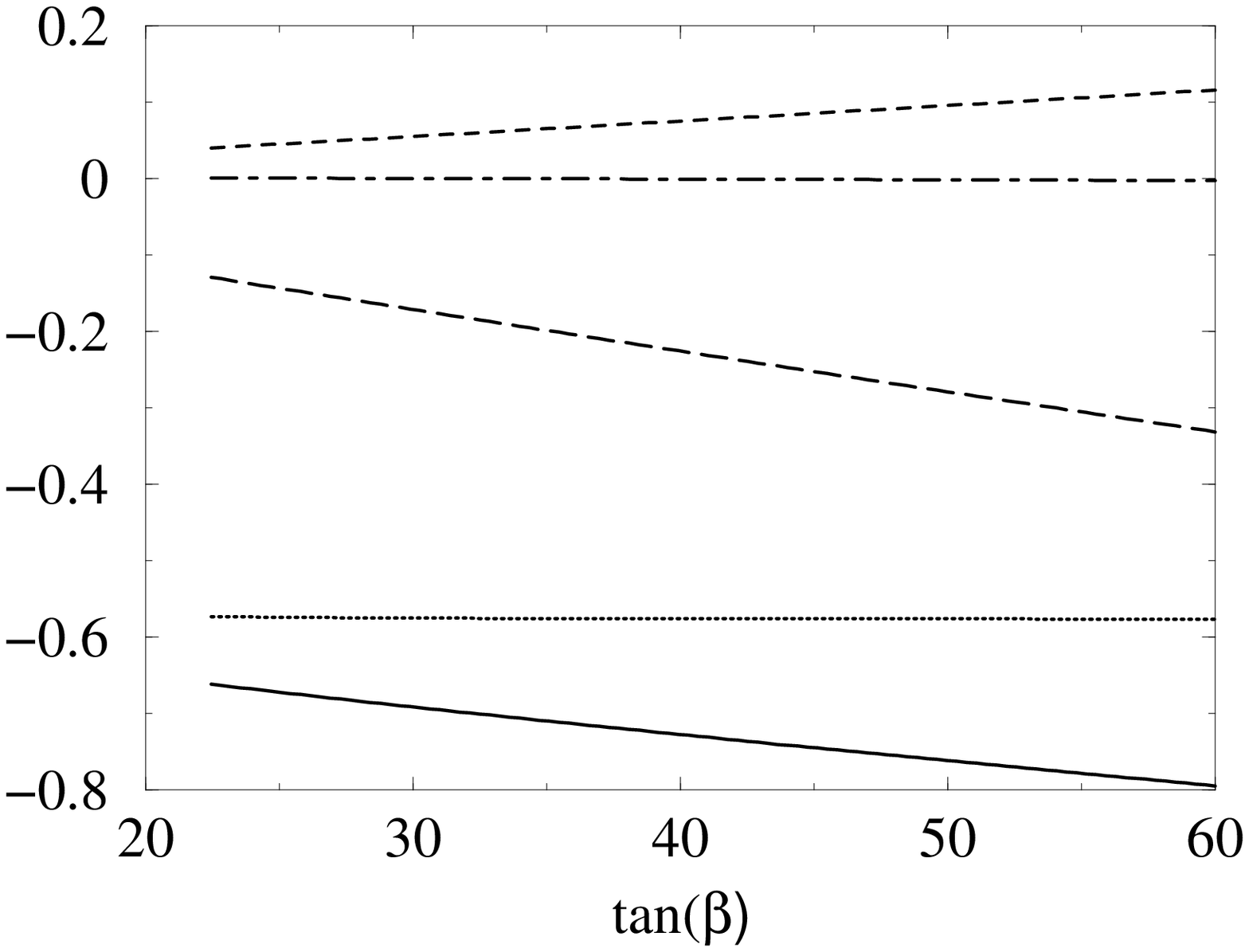,width=7cm} \\
(a) & (b)
\end{tabular}
\end{center}
\caption{The correction $\delta$, eq.\,(\ref{eq:defdelta}),
  $\Gamma(t \rightarrow H^+ b)$ as a function of $\tan\beta$,
  for the MSSM and for two different scenarios,
  {\bf (a)}  set A: \{\mHp=120, 
  $\mu=-90$, M=150, \msbo=150, \msto=100,
 $m_{\tilde u}=m_{\tilde \nu}$=400, \mg=300,
 $A_t=A_b=A_{up}=A_{l}$=300\} GeV
 {\bf (b)} set B: 
 \{\mHp=120,
$\mu=+90$, M=150 , \msbo=\msto=400,
 $m_{\tilde u}=m_{\tilde \nu}$=400, \mg=1000,
$A_t=-500$, $A_b=A_{up}=A_{l}$=300 \} GeV. Shown are: 
the Higgs sector
  contribution $\delta_{\rm Higgs}$; the contribution from 
  the supersymmetric
  electroweak sector $\delta{\rm SUSY-EW}$; the supersymmetric 
  QCD contribution
  $\delta_{QCD}$; the standard QCD contribution $\delta_{QCD}$; 
  and the total
  correction $\delta_{\rm Total}$.
}\label{fig:deltatbmssm}
\end{figure}
}

\newcommand{\figmasthdm}{
\begin{figure}[t]
\begin{center}
\begin{tabular}{cc}
\epsfig{file=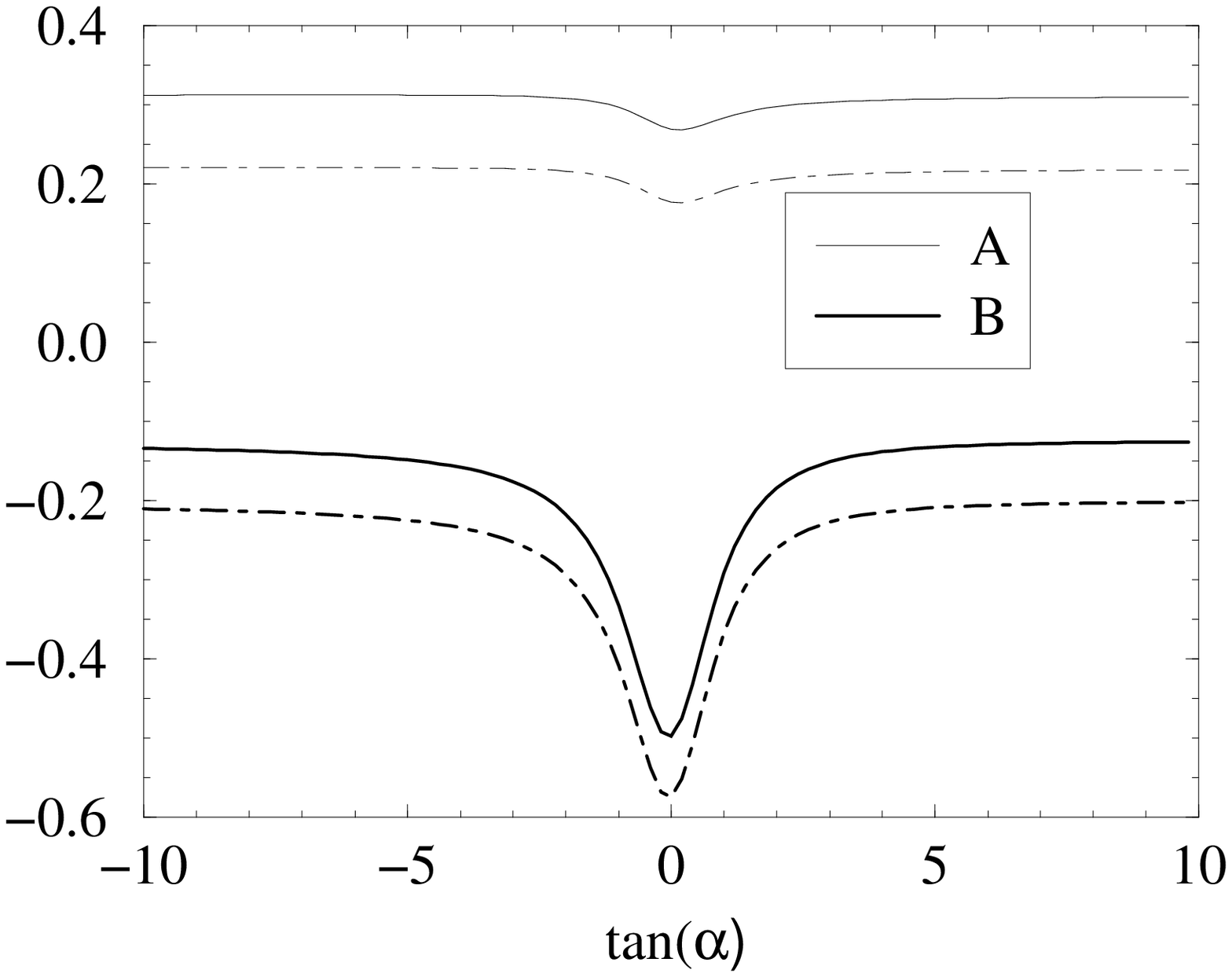,width=7cm} & 
 \epsfig{file=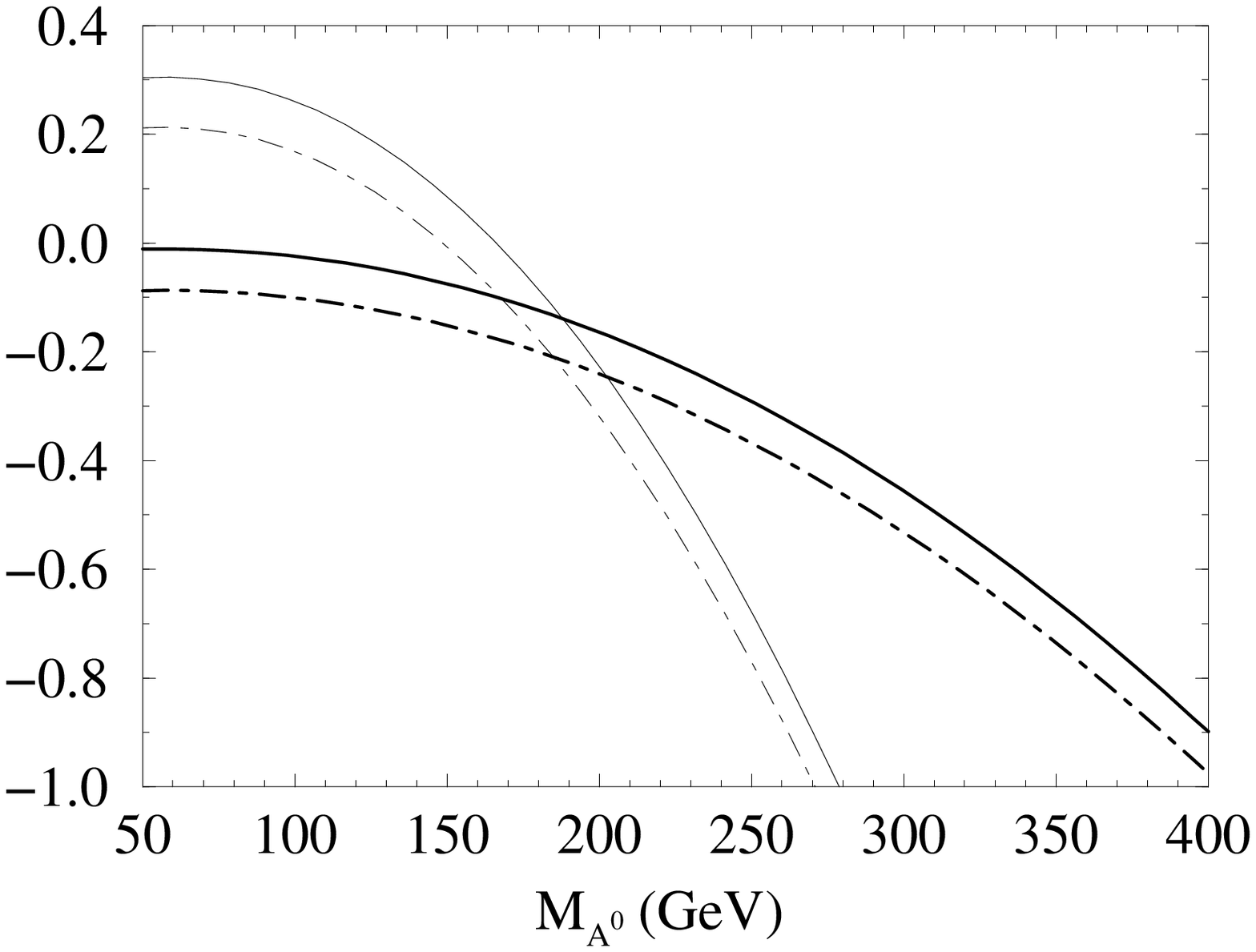,width=7cm} \\ 
{ (a)} & { (b)}\\  
\epsfig{file=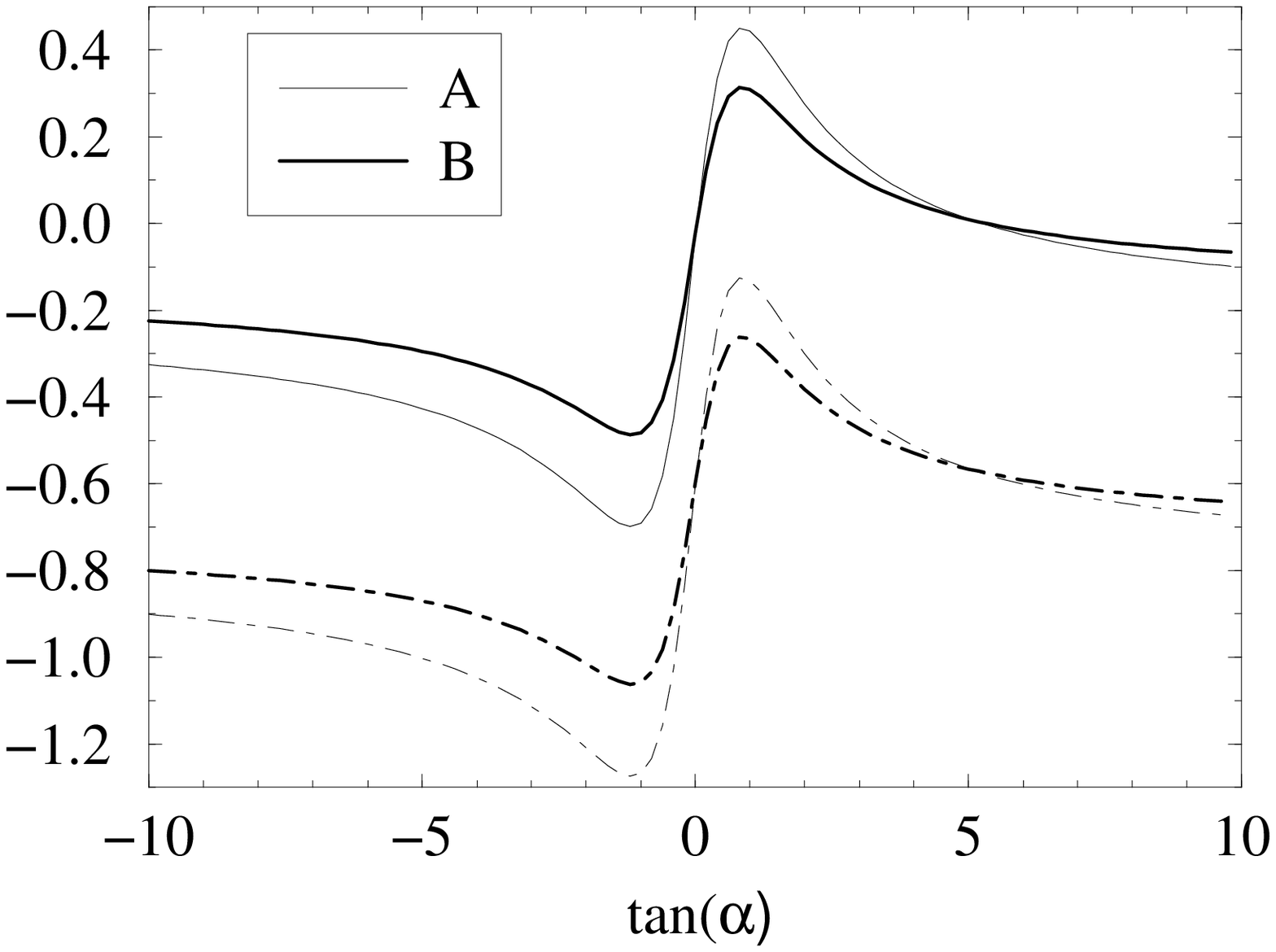,width=7cm} &
 \epsfig{file=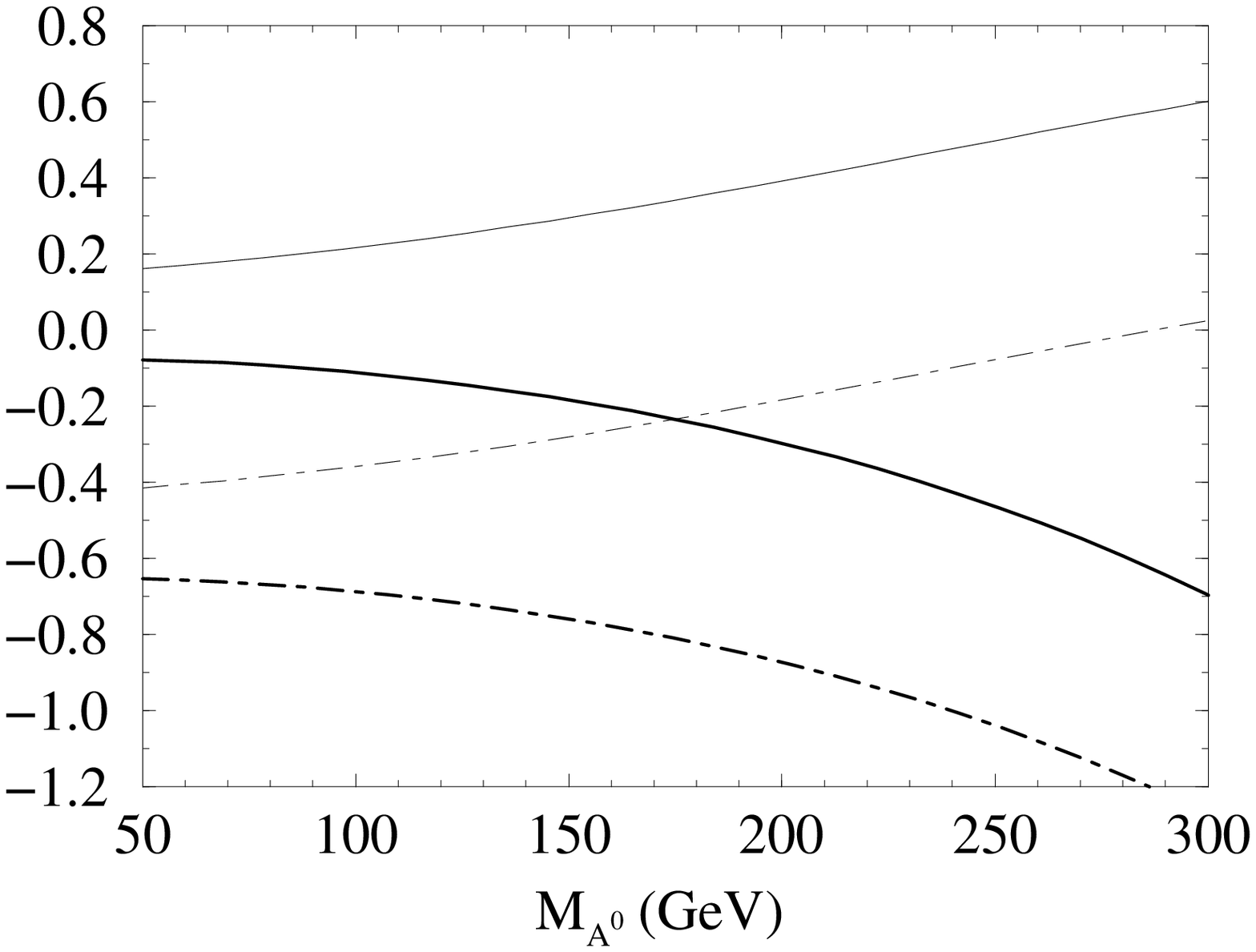,width=7cm} \\ 
 { (c)} & { (d)}
\end{tabular}
\end{center}
\caption{The corrections  $\delta_{\rm EW}$ and
$\delta_{\rm Total}$
  {\bf (a)} 
  for the Type I $2$HDM as a function of $\tan\alpha$, 
  inputs as in Fig.\,\ref{fig:deltatb2hdm} with $\tb=0.1$ 
  for set A and $\tb=0.2$ for set B,
  {\bf (b)} as in (a) but for the pseudoscalar Higgs mass,
  {\bf (c)} as in (a) but for Type II $2$HDM with $\tb=35$ 
  for both sets, 
  {\bf (d)} as in (c) but for the pseudoscalar Higgs mass.}%
  \label{fig:deltamass2hdmI}
\end{figure}
}

\newcommand{\figatmssm}{
\begin{figure}
\begin{center}
\epsfig{file=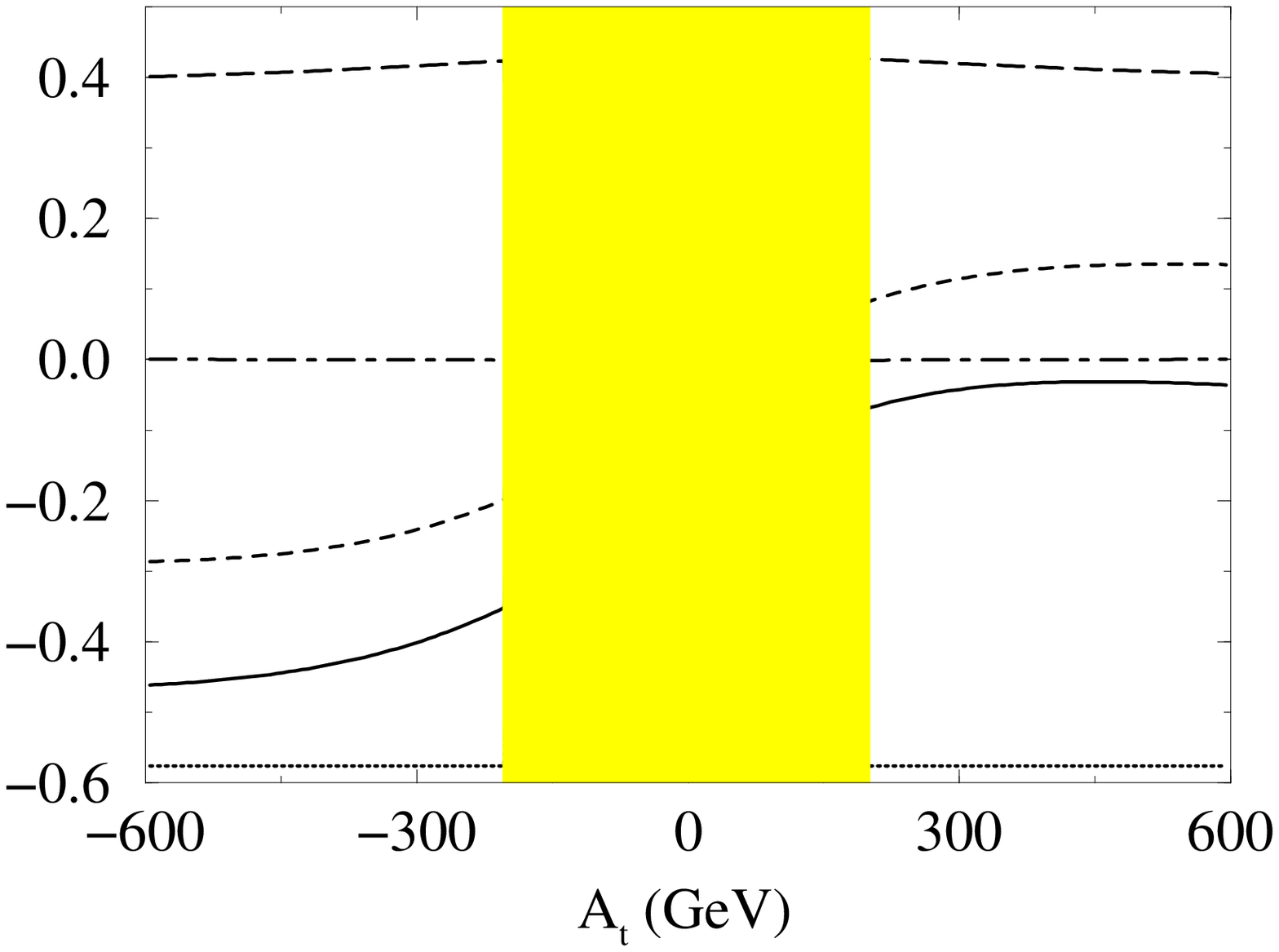,width=7cm}
\end{center}
\caption{The correction $\delta$ for the MSSM as a function of the
  Soft-SUSY-breaking trilinear coupling $A_t$. Inputs as in as in
  Fig.\,\ref{fig:deltatbmssm}(a). Shown are the same contributions 
  as in
  Fig.\,\ref{fig:deltatbmssm}.}\label{fig:deltaAtmssm}
\end{figure}
}

\newcommand{\figexcthdm}{
\begin{figure}
\begin{center}
\begin{tabular}{cc}
\epsfig{file=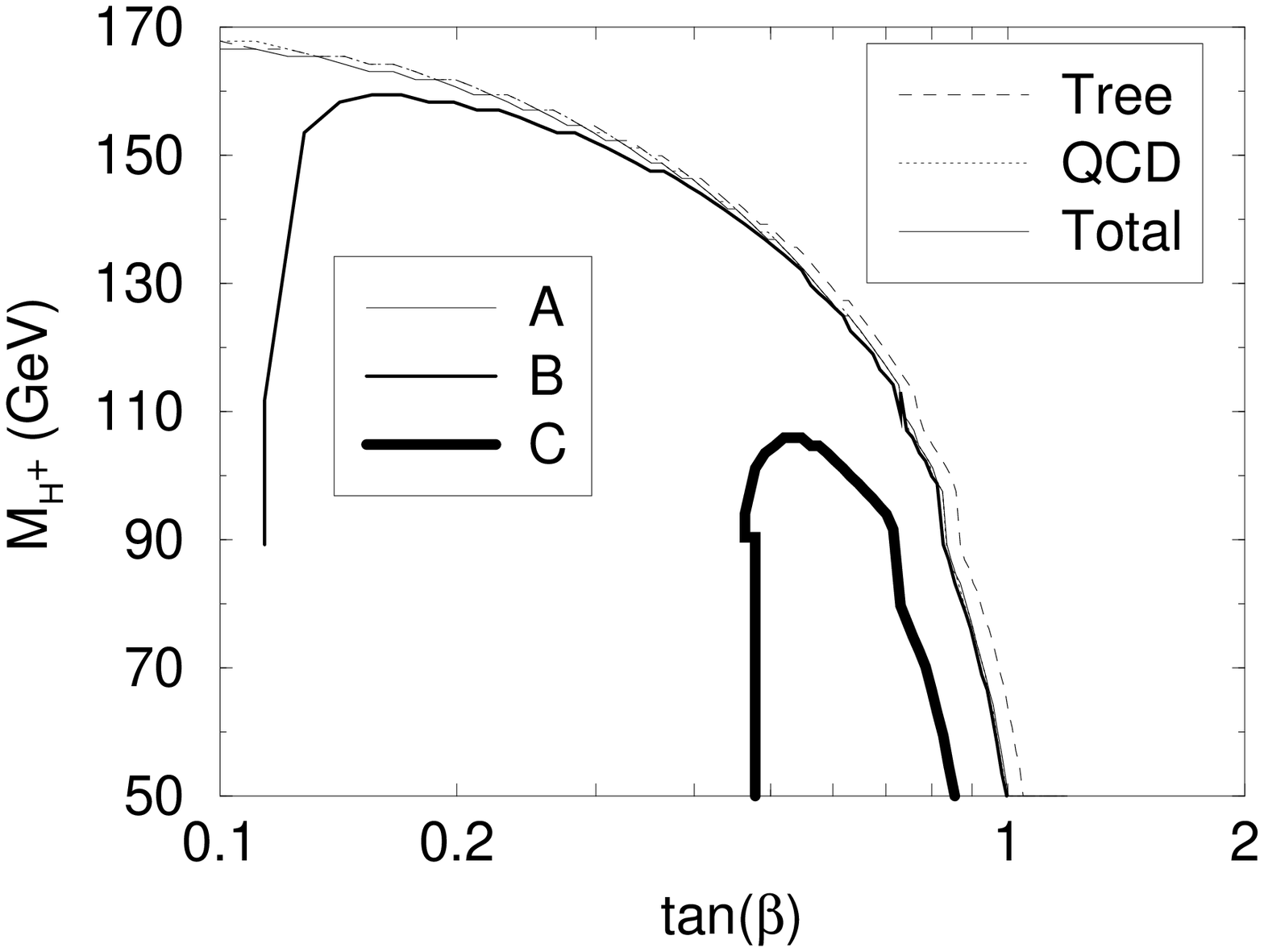,width=7cm} & 
 \epsfig{file=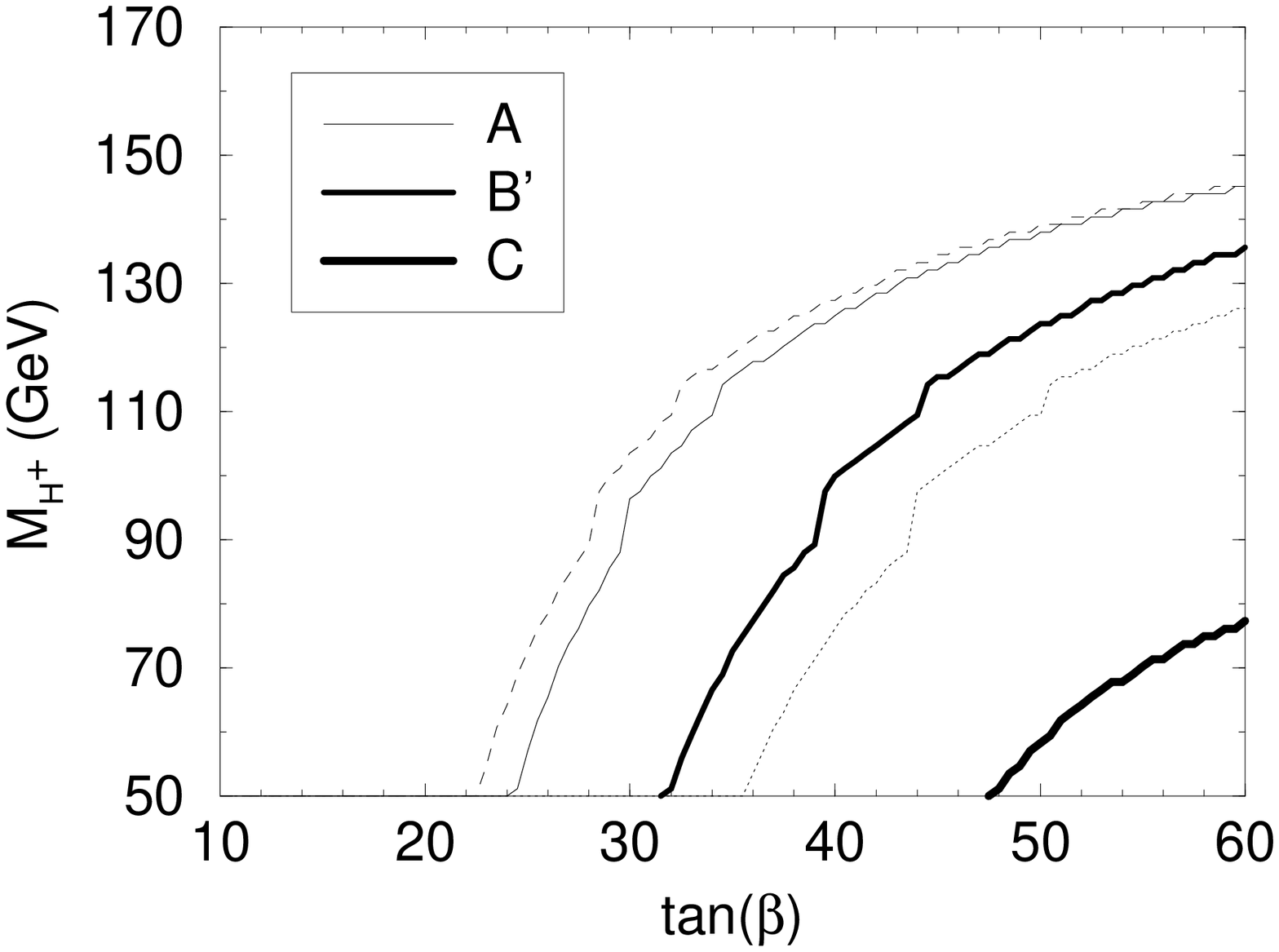,width=7cm}\\ 
{ (a)} & { (b)} \\
\end{tabular}
\end{center}
\caption{The $95\%$ C.L. exclusion plot in the
  $(\tb, \mh)$-plane for
  {\bf (a)} Type I $2$HDM using three sets of inputs:
  A and B as in Fig.\,\ref{fig:deltatb2hdm}, and
  C: $\{(\mh,200,80,700)\, GeV;\,1\}$;
  {\bf (b)} Similarly for Type II models including three sets 
  of inputs:
  A as defined in Fig.\,\ref{fig:deltatb2hdm},
  B':$\{(\mh,200,80,150)\,GeV;\,0.3\}$, and
  C:$\{(\mh,200,80,150)\,GeV;\,-3\}$.
  Shown are the tree-level, QCD-corrected and fully 2HDM-corrected
  contour lines.  The excluded region in each case
  is the one lying below these curves}\label{fig:exc2hdm}
\end{figure}
}
\newcommand{\figexcmssm}{
\begin{figure}
\begin{center}
\epsfig{file=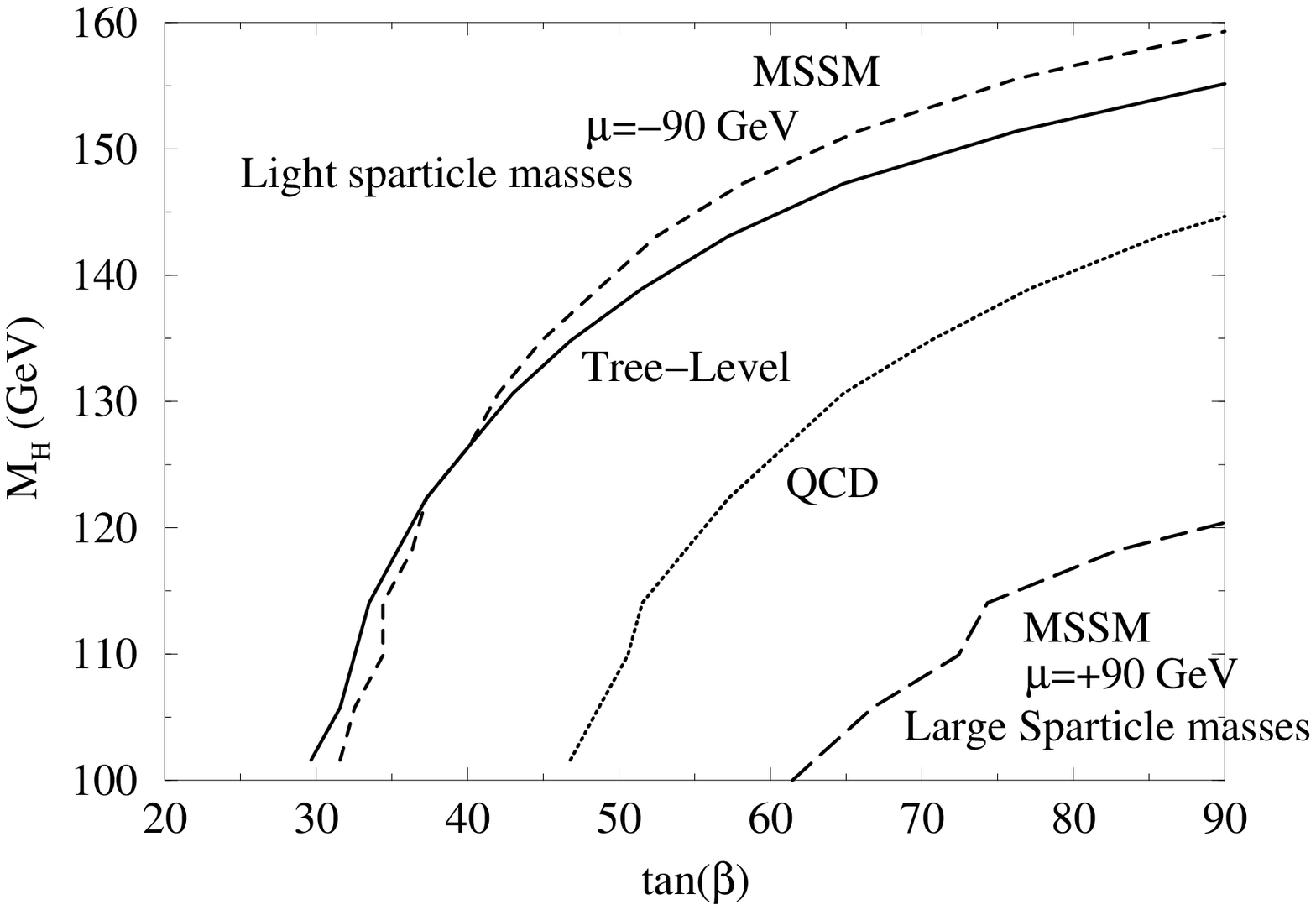,width=7cm}  
\end{center}
\caption{The $95\%$ C.L. exclusion plot in the
  $(\tb, \mh)$-plane for the MSSM. Shown are the excluded region 
  using: the
  tree-level prediction for $\Gamma(t\to H^+\,b)$; the standard 
  QCD prediction;
  and the full MSSM predictions for sets A and B as in 
  Fig.\,\ref{fig:deltatbmssm}.}\label{fig:excmssm}
\end{figure}
}

\begin{document}
\def\pubnum{451}
\def\data{February, 1999}

\def\UUAABB{\hbox{
    \vrule height0pt width2.5in
    \vbox{\hbox{\rm 
     UAB-FT-\pubnum
    }\break\hbox{\data\hfill}
    \break\hbox{hep-ph/9903212\hfill} 
    \hrule height2.7cm width0pt}
   }}   
\hfill\UUAABB
\vspace{3cm}
\begin{center}
\begin{large}
\begin{bf}
RADIATIVE CORRECTIONS TO TOP QUARK DECAY INTO CHARGED 
HIGGS AT THE TEVATRON\footnote{Talk presented
 at the IVth International Symposium on Radiative 
 Corrections (RADCOR 98),
 Barcelona, September 8-12, 1998. To appear in 
 the proceedings, World Scientific, ed. J. Sol{\`a}.}\\
\end{bf}
\end{large}
\vspace{1cm}
J.A. COARASA, Jaume GUASCH, Joan SOL{\`A}\\

\vspace{0.25cm} 
Grup de F{\'\i}sica Te{\`o}rica\\ 
and\\ 
Institut de F{\'\i}sica d'Altes Energies\\ 
\vspace{0.25cm} 
Universitat Aut{\`o}noma de Barcelona\\
08193 Bellaterra (Barcelona), Catalonia, Spain\\
\end{center}
\vspace{0.3cm}
\hyphenation{super-symme-tric sig-ni-fi-cant-ly ge-ne-ral
asso-cia-ted re-mar-ka-bly}
\hyphenation{com-pe-ti-ti-ve}
\hyphenation{mo-dels}
\begin{center}
{\bf ABSTRACT}
\end{center}
\begin{quotation}
\noindent
\hyphenation{ob-ser-va-bles com-pa-ti-bi-li-ty}
\noindent
We present the computation of the leading one-loop
  electroweak radiative corrections to the non-standard 
  top quark decay width
  $\Gamma(t\rightarrow H^+\,b)$, using a physically 
  motivated definition of
  \tb. We find that the corrections are large, both in the Minimal
  Supersymmetric Standard Model (MSSM) and the 
  Two-Higgs-Doublet Model
  (2HDM). These corrections have an important effect 
  on the interpretation of
  the Tevatron data, leading to the non-existence of 
  a model-independent bound
  in the $\tb-\mHp$ plane.
\end{quotation}
 
\baselineskip=6.5mm  
 
\newpage

\section{Introduction}
The top quark has been subject of dedicated studies since its 
discovery at the Fermilab Tevatron Collider\cite{Tevatron}.  Due to 
its large mass it can develop large couplings with the Spontaneus 
Symmetry Breaking sector of the theory, and the Electroweak quantum 
corrections of this sector could be large, and indeed they are.  
This 
is specially true in some extensions of the Standard Model (SM) 
where 
this sector is enlarged, such as the Two-Higgs-Doublet Model 
(2HDM)\cite{Hunter}, or the Minimal Supersymmetric Standard Model 
(MSSM).

Here we will present the computation of the Electroweak corrections 
to 
the non-standard top quark decay partial width into a charged Higgs 
particle and a bottom quark $\Gamma(t\to H^+\,b)$.  We will present 
the correccions arising in generic Type I and Type II 2HDM, as well 
as 
the MSSM.  We make our computation at leading order in both the 
Yukawa 
coupling of the top quark, and the Yukawa coupling of the bottom 
quark.

The Two-Higgs-Doublet Model  ($2$HDM)\cite{Hunter} plays a special 
role
as the simplest extension of the electroweak sector of
the SM. 
After spontaneous symmetry breaking one is left with two CP-even
(scalar) Higgs bosons $h^0$, $H^0$, a CP-odd
(pseudoscalar) Higgs boson $A^0$ and a pair of charged Higgs
bosons $H^\pm$.  The parameters of these models consist of: i) the
masses of the Higgs particles,
$M_{h^0}$, $M_{H^0}$, $M_{A^0}$ and $\mh$
(with the convention $M_{h^0}<M_{H^0}$),
ii) the
ratio of the two {vacuum expectation values}: $\tb={v_2 / v_1}$,
and the mixing angle $\alpha$ between the two CP-even states.
Two types of such models have been of special interest\cite{Hunter} 
which
avoid potentially 
dangerous tree-level
Flavour Changing Neutral Currents: In Type I $2$HDM only one of the
Higgs doublets is coupled to the fermionic sector,
whereas in Type II $2$HDM each Higgs doublet ($H_1$, $H_2$) is 
coupled
to the up-type fermions and down-type fermions respectively, the
Yukawa couplings being
\beq
\lambda_t\equiv {h_t\over g}={m_t\over \sqrt{2}\,M_W\,
\sin{\beta}}\;\;\;\;\;,
\;\;\;\;\; \lambda_b^{\{{\rm I,\,II}\}}\equiv {h_b\over g}=
{m_b\over \sqrt{2}
\,M_W\,\{\sin{\beta},\cos{\beta}\}}\,\,.
\label{eq:Yukawas}
\eeq
Type II models do appear in specific
extensions of the SM,
such as the Minimal Supersymmetric Standard Model (MSSM) which is
currently under intensive study both theoretically and 
experimentally. In this
latter model all the parameters of the Higgs sector are 
computed as a function
of just two input parameters: $\tb$ and a mass, which 
we take to be
$\mh$\,\footnote{We use the one-loop MSSM Higgs bosons 
mass relations\cite{Dabelstein} to compute the rest of 
the masses and $\tan\alpha$.}.

In case that
the charged Higgs boson is light enough, the top
quark  could decay via the non-standard channel 
$t\rightarrow H^+\, b$.  Based
on this possibility the CDF collaboration at the Tevatron
has undertaken
an experimental program which at the moment has been used 
to put limits on 
the parameter space of Type II models\cite{CDF}.
The bounds are
obtained by searching for an excess of the cross-section
$\sigma(p\bar{p}\rightarrow t \bar{t} X\rightarrow 
\tau \nu_{\tau} X)$ with
respect to
$\sigma(p\bar{p}\rightarrow t\bar{t}X\rightarrow
l\nu_{l}\,X)$ ($l= e,\,\mu$).
The absence of such an excess determines an upper bound on
$\Gamma(t\rightarrow H^+\,b\rightarrow\tau^+\,\nu_\tau\,b)$
and a corresponding
excluded region of the parameter space $(\tb,\mh)$.
However, it has  been shown that the one-loop quantum corrections 
to that
decay width can be rather large. This applies not only to the
conventional QCD one-loop corrections\cite{CD}
-- the only ones used in Ref.\cite{CDF} --
but also to the QCD and electroweak corrections in the framework 
of the
MSSM\cite{GJS,CGGJS,TesiJaume}. Thus the CDF limits could be 
substantially modified by
radiative  corrections\cite{GuaschSola} and in some cases the bound
even disappears.

We remark that although CLEO data on $BR(b\rightarrow s \gamma)$ 
could
preclude the existence of a light charged Higgs
boson~\cite{CLEO} -- thus barring the possibility of the top quark
 decaying into it -- this assertion is not completely general and, 
 moreover,
needs
further experimental confirmation\footnote{See Ref.\cite{CGSH} for 
details.}.

It is our aim to investigate, independent of and complementary to
the indirect constraints,
the decay $t\rightarrow H^+\,b$ in general $2$HDM's (Types I and
II) and in the MSSM by strictly taking into consideration the 
direct data from
Tevatron on equal footing as in Ref.\cite{CDF}.
This study should be useful to 
distinguish the kind of quantum effects expected in general 
$2$HDM's as
compared to those within the context of the MSSM.

The interaction Lagrangian describing the
$H\,t\,b$-vertex  in Type-$j$ $2$HDM $(j=I,II)$ is:
\beq
{\cal L}_{H t b}^{(j)}=\frac{g}{\sqrt{2}M_W} H^-\,
 \bar{b}\, \left[m_t\, \ctb\,\pr +
 m_b\, a_j\, \pl \right]\, t + {\rm h.c.}
\label{eq:interaction}
\eeq
where we have introduced the parameter $a_j$ with $a_I\equiv -\ctb$,
$a_{II}\equiv +\tb$.  From the
interaction Lagrangian~(\ref{eq:interaction}) it is patent that
for Type I models the
branching ratios $BR(t\rightarrow H^+\,b)$ and $BR(H^+\rightarrow
\tau^+ \nu_\tau)$ are relevant only at low $\tb$,
whereas for Type II models the former branching ratio can be 
important both at
low and high $\tb$ and the latter is only significant at high
values of $\tb$.

\section{One-loop corrected $\Gamma(t\to H^+\,b)$}
The renormalization procedure
required for the one-loop amplitude extends that of 
Ref.~\cite{CGGJS}.
The counterterm Lagrangian $\delta{\cal L}_{Hbt}^{(j)}$ for each
$2$HDM model $j=I,II$ reads
\beq
\delta{\cal L}_{Hbt}^{(j)}={g\over\sqrt{2}\,M_W}\,H^-\,\bar{b}
\left[
\delta C_R^{(j)}\ m_t\,\ctb\,\,P_R+
\delta C_L^{(j)}\ m_b\,a_j\,P_L\right]\,t
+{\rm h.c.}\,,
\label{eq:LtbH2}
\eeq
with
\beqn
\delta C_R^{(j)} &=& {\delta m_t\over m_t}-{\delta v\over v}
+\frac{1}{2}\,\delta Z_{H^+}+\frac{1}{2}\,\delta Z_L^b+\frac{1}{2}
\,\delta Z_R^t
-{\delta\tb\over\tb}+\delta Z_{HW}\,\tb\,,\nonumber\\
\delta C_L^{(j)} &=& {\delta m_b\over m_b}-{\delta v\over v}
+\frac{1}{2}\,\delta Z_{H^+}+\frac{1}{2}\,\delta Z_L^t+\frac{1}{2}
\,\delta Z_R^b
\mp{\delta\tb\over\tb}\,-\delta Z_{HW}\,\frac{1}{a_j}\,,
\label{eq:deltacgen}
\eeqn
where in the last expression the upper minus sign applies to Type I 
models 
and the lower plus sign to Type II -- hereafter we will adopt
this convention.

The counterterm $\delta\tb/\tb$ is defined in such a way that
it absorbs the one-loop contribution to
 the decay
width $\Gamma(H^+\rightarrow \tau^+ \nu_\tau)$,
yielding
\beq
{\delta\tb\over \tb}
=\mp\left[{\delta v\over v}-\frac{1}{2}\delta Z_{H^\pm}
+ \delta Z_{HW}\frac{1}{a_j}+
\Delta_{\tau}^{(j)}\,\right]\,\,.
\label{eq:deltabeta}
\eeq
The quantity
\beq
\Delta_{\tau}^{(j)}=-{\delta m_{\tau}\over m_{\tau}} -
\frac{1}{2}\delta
Z_L^{\nu_{\tau}}-\frac{1}{2}\delta Z_R^{\tau}-F_{\tau}^{(j)}\,,
\label{eq:deltatau}
\eeq
contains the (finite) process-dependent part of the counterterm, 
where
$F_\tau$
comprises the complete set of one-particle-irreducible three-point 
functions
of the charged Higgs decay into $\tau^+\,\nu_\tau$.

The correction to the decay width in each $2$HDM
is defined as
\beq
\delta_{\rm 2HDM}^{(j)} = \frac{\Gamma^{(j)}(t\rightarrow
 H^+\,b)-\Gamma_0^{(j)}(t\rightarrow H^+\,b)}
{\Gamma_0^{(j)}(t\rightarrow
 H^+\,b)}
\label{eq:defdelta}
\eeq
where $\Gamma_0^{(j)}$ is the lowest-order width
in the on-shell $\alpha$-scheme\,\footnote{See Ref.\cite{CGSH}.}.

The renormalized one-loop vertices $\Lambda_{L,R}$
for each type of model are obtained
after adding up the
counterterms~(\ref{eq:deltacgen}) to the one-loop form factors:
\beqn
\Lambda_L&=&\delta C_L + F_L\nonumber\\
\Lambda_R&=&\delta C_R + F_R\,\,\,.
\eeqn
The one-loop Feynman diagrams contributing to the decay
$t \rightarrow H^+ b$ under consideration can be seen in 
Ref.~\cite{CGGJS}. For
the $2$HDM one just must take the Higgs bosons mediated diagrams:
Fig.\,3 (all diagrams), Fig.\,4 (diagrams $C_{b3}$,
$C_{b4}$, $C_{t3}$, $C_{t4}$), Fig.\,5 (diagram $C_{H1}$) and 
Fig.\,6 (diagram
$C_{M1}$) of that reference. It goes without saying that the 
calculation of
these diagrams in general $2$HDM's is different from that in
Ref.\cite{CGGJS},
and this is so even for the Type II case since some of the
Higgs boson Feynman rules for supersymmetric models\cite{Hunter} 
cannot
be borrowed without a careful adaptation of the
couplings\,\footnote{We have generated a
fully consistent set. In part they can be found in \cite{HollikZP} 
and
references therein. See also\cite{TesiToni}.}. 

\subsection{Vertex functions}

Now we consider contributions arising from the exchange of virtual
Higgs particles
and Goldstone bosons in the Feynman gauge,
as shown in Fig.3 of Ref.\,\cite{CGGJS}. We follow the vertex
formula for the form factors by the value of the
overall coefficient $N$ 
and by the arguments of the corresponding $3$-point functions.

We start by defining the following factors for each Type-$j$ 2HDM:
\begin{eqnarray}
R_{j}&=&\{ \sa /\sbt , \ca / \cbt  \}\,\,,\nonumber\\
r_{j}&=&\{ \ca /\sbt ,  - \sa / \cbt  \}\,\,,\nonumber
\end{eqnarray}
then the contributions from the different diagrams can be written as
\begin{itemize}
\item {Diagram $(V_{H1})$:}
\begin{eqnarray}
  F_L&=&N\,[\mbs (\Cot-\Cz)+\mts \ctbsH (\Coo-\Cot) ]\,,\nonumber\\
  F_R&=&N\mbs [\Cot-\Cz +\tbsH (\Coo-\Cot) ] \,,\nonumber\\
  N&=&-\frac{ig^2}{2}
        \{R_{j} , r_{j} \} N_{1}\, ,\nonumber \\
  N_{1}&=& \frac{\mAs-M_{\{H^{0},h^{0}\}}^{2}}{\mws}
  \cot\!2\beta \{\sab ,\cab\}+ \nonumber \\
&& \frac{\mAs-\mHp^{2}-M_{\{H^{0},h^{0}\}}^{2}/2}{\mws} 
\{\cbma ,\sbma\}\,,
\label{eq:vh1}
\end{eqnarray}
where $C_{*}$ are the usual one-loop scalar three-point
functions\,\cite{Veltmann}. In eq.\,(\ref{eq:vh1}) they must 
be evaluated with
arguments
$$
  C_{*}=C_{*}\left(p,p',\mb,\mHp,\{\mH,\mhz\}\right)\,.
$$
\item {Diagram $(V_{H2})$:}
\begin{eqnarray*}
  F_L&=&N\frac{1}{a_{j}}[\mts (\Coo -\Cot )+\mbs(\Cz -\Cot )]\,,\\
  F_R&=&N\mbs\tb (2\Cot -\Coo -\Cz ) \,,\\
  N&=&\pm\frac{ig^2}{4} \{R_{j} , r_{j} \}\{\sbma  , \cbma \}
        \left(\frac{\mHps}{\mws}-\frac{\{\mHs  , \mhzs \}}
{\mws}\right)\,, \\
  C_{*}&=&C_{*}\left(p,p',\mb,\mw,\{\mH,\mhz\}\right)\,.
\end{eqnarray*}

\item {Diagram $(V_{H3})$:}
\begin{eqnarray*}
  F_L&=&N\mts [\ctbsH\Cot +\Coo -\Cot -\Cz ] \,,\\
  F_R&=&N\,[\mbs\tbsH\Cot  +\mts (\Coo -\Cot-\Cz )]\,, \\
  N&=&-\frac{ig^2}{2}\frac{\{\sa , \ca \}}{\sbt} N_{1}\,,\\
  C_{*}&=&C_{*}\left(p,p',\mt,\{\mH,\mhz\},\mHp\right)\,.
\end{eqnarray*}

\item {Diagram $(V_{H4})$:}
\begin{eqnarray*}
  F_L&=&N\mts (2\Cot-\Coo+\Cz)\frac{1}{a_{j}} \,,\\
  F_R&=&N\,[-\mbs\Cot +\mts (\Coo-\Cot-\Cz)]\tb\,, \\
  N&=&\mp\frac{ig^2}{4}\frac{\{\sa \sbma, \ca \cbma\}}{\sbt} %
        \left(\frac{\mHps}{\mws}-\frac{\{\mHs  ,\mhzs \}}
{\mws}\right) \,,\\
  C_{*}&=&C_{*}\left(p,p',\mt,\{\mH,\mhz\},\mw\right)\,.
\end{eqnarray*}

\item {Diagram $(V_{H5})$:}
\begin{eqnarray*}
  F_L&=&N\,[\mbs(\Cot+\Cz)+\mts(\Coo-\Cot)]\,, \\
  F_R&=&N \mbs \tbsH (\Coo+\Cz)\,, \\
  N&=&-\frac{ig^2}{4}\left(\frac{\mHps}{\mws}-\frac{\mAs}
{\mws}\right)\,, \\
  C_{*}&=&C_{*}\left(p,p',\mb,\mw,\mA\right)\,.
\end{eqnarray*}

\item {Diagram $(V_{H6})$:}
\begin{eqnarray*}
  F_L&=& N\mts\ctbsH (\Coo+\Cz)\,, \\
  F_R&=& N\,[\mbs\Cot +\mts (\Coo-\Cot+\Cz)]\,, \\
  N&=&-\frac{ig^2}{4}\left(\frac{\mHps}{\mws}-\frac{\mAs}
{\mws}\right) \,,\\
  C_{*}&=&C_{*}\left(p,p',\mt,\mA,\mw\right)\,.
\end{eqnarray*}

\item {Diagram $(V_{H7})$:}
\begin{eqnarray*}
  F_L&=&N\,[(2\mbs\Coo+\Czt+2(\mts-\mbs)(\Coo-\Cot ))\ctbsH 
  \nonumber\\
&&        +2\mbs(\Coo+2\Cz)]\mts\,, \\
  F_R&=&N\,[(2\mbs\Coo+\Czt+2(\mts-\mbs)(\Coo-\Cot ))\tbsH 
  \nonumber\\
&&        +2\mts(\Coo+2\Cz)]\mbs\,, \\
  N&=&\frac{ig^2}{4\mws}\left\{ \frac{\sa}{\sbt}  R_{j}\, ,
 \frac{\ca}{\sbt} r_{j}  \right\} \,, \\
  C_{*}&=&C_{*}\left(p,p',\{\mH,\mhz\},\mt,\mb\right)\,.
\end{eqnarray*}

\item {Diagram $(V_{H8})$:}
\begin{eqnarray*}
  F_L&=& N\mts\{\ctbsH \, ,\ctbs \} \,\Czt\,,  \\
  F_R&=& N\mbs\{\tbsH \, , \tbs\} \,\Czt\,,   \\
  N&=&\mp \frac{ig^2}{4\mws}\,, \\
  C_{*}&=&C_{*}\left(p,p',\{\mA,\mz\},\mt,\mb\right)\,.
\end{eqnarray*}

\end{itemize}

\subsection{Counterterms}
The diagrammatic contributions to the various counterterms in
eq.\,(\ref{eq:deltacgen}) can bee seen in Ref.\,\cite{CGGJS}: 
in Fig.\,4 (diagrams $C_{b3}$,
$C_{b4}$, $C_{t3}$, $C_{t4}$) the external fermion counterterms; 
In Fig.\,5
(diagram $C_{H1}$) the charged Higgs particle counterterms; 
And in Fig.\,6
(diagram $C_{M1}$) the contribution to the $W^\pm-H^\pm$ mixing 
self-energy
contributions. Now we pass to the explicit expression of that 
counterterms.
\begin{itemize}
\item {Counterterms $\delta m_f\,,\delta Z_L^f\,,\delta Z_R^f$:}
For a given down-like fermion $b$, and corresponding
isospin partner $t$, 
the fermionic self-energies receive contributions
\begin{eqnarray}
  \label{HiggsSelfb}
  \Sigma_{\{L,R\}}^b(p^2)&=&\left.\Sigma_{\{L,R\}}^b(p^2)
   \right|_{(C_{b3})+(C_{b4})}=\frac{g^2}{2i\mws}\nonumber\\
  &\times&\left\{m_{\{t,b\}}^2\left[
   \{\ctbs,a_{j}^{2}\} \Bo(p,\mt,\mHp)+\Bo(p,\mt,\mw)
   \right]\right. \nonumber\\
  &+&\frac{\mbs}{2}\left[\casH\,\Bo(p,\mb,\mH)
   +\sasH\,\Bo(p,\mb,\mhz)\right. \nonumber\\
  &&\ \ \ +\left.\left. a_{j}^{2}\,\Bo(p,\mb,\mA)
   +\Bo(p,\mb,\mz)\right]\right\}\,,\nonumber\\
  \Sigma_S^b(p^2)&=&\left.\Sigma_S^b(p^2)
   \right|_{(C_{b3})+(C_{b4})}\nonumber\\
 &=&-\frac{g^2}{2i\mws}\left\{\mts a_{j} \ctb \left[
   \Bz(p,\mt,\mHp)-\Bz(p,\mt,\mw)\right]\right.\nonumber\\
  &+&\frac{\mbs}{2}\left[\casH\,\Bz(p,\mb,\mH)+
   \sasH\,\Bz(p,\mb,\mhz)\right.\nonumber\\
  &&\ \ \ -\left.\left. a_{j}^{2}\,\Bz(p,\mb,\mA)
   -\,\Bz(p,\mb,\mz)\right]\right\}\,,
\end{eqnarray}
from Higgs and Goldstone bosons in the Feynman gauge.
To obtain the corresponding expressions for an up-like fermion, 
$t$, just
perform the label substitutions $b\leftrightarrow t$ and 
replace $R_j\to \sa/\sbt$,  $r_j\to \ca/\sbt$, 
$a_j \leftrightarrow \ctb$ on eq.~(\ref{HiggsSelfb}).


Now one must introduce that expressions into the standard 
on-shell definitions
of $\delta m_f$ and $\delta Z_{L,R}^f$ (see e.g. eqs.(20) 
and (21) of
Ref.\,\cite{CGGJS}).
\item {Counterterm $\delta Z_{H^\pm}$}:
\begin{eqnarray}
  \label{dZHSusy}
  \delta Z_{H^\pm}&=&\left.\delta Z_{H^\pm}
                     \right|_{(C_{H1})}
    =\,\Sigma_{H^\pm}'(\mHps)\nonumber\\
  &=&-\frac{ig^2N_C}{\mws}\left[(\mbs a_{j}^{2}+\mts\ctbs)
   (\Bo+\mHps\Bo'+\mbs\Bz')\right. \nonumber\\
  &&\left.\mbox{\hspace{1.5cm}}+
  2\ctb a_{j}\mbs\mts\Bz'\right](\mHp,\mb,\mt)\,.
\end{eqnarray}
\item {Counterterm $\delta Z_{HW}$}:
\begin{eqnarray}
  \label{dZHWSusy}
  \delta Z_{HW}&=&\left.\delta Z_{HW}\right|_{(C_{M1})}
 =\frac{\Sigma_{HW}(\mHps)}{\mws}\nonumber\\
 =&-&\frac{ig^2N_C}{\mws}\left[\mbs a_{j}(\Bz+\Bo)
  +\mts\ctb\Bo\right](\mHp,\mb,\mt)\,. 
\end{eqnarray}
A sum is understood over all generations.
\end{itemize}


\section{Numerical analysis}
In the numerical analysis presented in
Figs.\,\ref{fig:deltatb2hdm}-\ref{fig:excmssm} we have put 
several cuts on 
our set of inputs\cite{CGSH}.
For $\tan\beta$ we have restricted in principle to the segment
\beq
0.1\stackm\tb\stackm 60\,\,\,.
\eeq
For the three Higgs bosons coupling we have imposed that they 
do not exceed
the maximum unitarity level permitted for the SM three Higgs 
boson coupling,
i.e.\footnote{We have corrected a misprint in eq.\,(16) of 
Ref.\cite{CGSH}.}
\beq
|\lambda_{HHH}|\stackm|\lambda_{HHH}^{SM}(m_H=
1\,TeV)|=g\frac{3}{2}\frac{(1\,TeV)^2}{M_W}\,\,.
\label{eq:hhh}
\eeq
This condition restricts both the ranges of masses and
of $\tan\beta$.
Moreover, we have imposed that the extra induced contributions to
the $\rho$ parameter are bounded by the
current experimental limit\,\footnote{Notice that this condition 
restrains 
$\Delta r$ within the experimental range and {\it a fortiori} the
corresponding corrections in the $G_F$-scheme. The bulk of
the EW effects are contained in the  non-universal corrections
predicted in the $\alpha$-scheme.} :
\beq
|\Delta\rho|\leq0.003\,\,.
\label{eq:deltarho}
\eeq
Of course in the MSSM analysis we apply all current limits on 
the SUSY
particles masses and parameters.

Before exploring the implications for the Tevatron analyses, 
we wish to show
the great sensitivity (through quantum effects) of the decay
$t\rightarrow H^+\,b$ to the particular structure of
the underlying $2$HDM.
In all cases we present our results in a significant region 
of the parameter
space where the branching ratios $BR(t\rightarrow H^+\,b)$ and
$BR(H^+\rightarrow \tau^+\,\nu_\tau)$ are expected to be sizeable. 
This
entails relatively light charged Higgs bosons 
($\mh\stackm 150\,GeV$) and a low
(high)  value of $\tb$ for Type I (II) models.

\figtbthdm

In Fig.\,\ref{fig:deltatb2hdm}  we display the evolution of the
correction~(\ref{eq:defdelta}) with $\tb$ for Types I and II 
$2$HDM's
and for two sets of inputs A and B for each model. We separately
show the (leading) EW contribution, $\delta_{\rm EW}$,
and the total correction, $\delta_{\rm Total}\equiv
\delta_{\rm EW}+\delta_{\rm QCD}$, which incorporates
the conventional QCD effects\cite{CD}. 
In the relevant $\tan\beta$
segments, that is below and above the uninteresting one,
we find that the pure EW contributions can be rather
large, to wit: For Type I models, the positive effects can 
reach $\simeq 30\%$,
while the negative contributions may increase `arbitrarily' --
thus effectively enhancing to a great extent the modest QCD 
corrections--
still in a region of parameter space respecting the various imposed
restrictions; For Type II models, instead, 
the EW effects can be very large, for both signs, in the high
$\tan\beta$ regime. In particular, the huge positive yields
could go into a complete ``screening'' of the QCD corrections.

\figtbmssm

In Fig.\,\ref{fig:deltatbmssm} we present the partial and total 
corrections in
the case of the MSSM. We present separately: the standard QCD 
corrections; the
supersymmetric (gluino-mediated) QCD correction\cite{GJS}; 
the Higgs boson
contributions; the supersymmetric contributions from the 
electroweak
sector\cite{CGGJS}; and the total correction, namely the net 
sum of all of the
above contributions. In Fig.\,\ref{fig:deltatbmssm}a we present 
and scenario
with $\mu<0$, and a relatively light sparticle spectrum. In
Fig.\,\ref{fig:deltatbmssm}b an scenario with $\mu>0$ and a heavy 
mass spectrum
is presented. The leading contribution to the MSSM correction is 
the bottom
quark mass finite threshold corrections  
--see eq.(\ref{eq:deltacgen})-- which reads\cite{CGGJS}
\begin{eqnarray}
\left({\delta m_b\over m_b}\right)_{%
\scriptscriptstyle{\rm SUSY-QCD}} &=&
 -{2\alpha_s(m_t)\over 3\pi}\,m_{\tilde{g}}{\mu\,}{\tan\beta}\,
I(m_{\tilde{b}_1},m_{\tilde{b}_2},m_{\tilde{g}}) \,,\nonumber\\
\left({\delta m_b\over m_b}\right)_{%
\scriptscriptstyle{\rm SUSY-Yukawa}} &=&
-{h_t^2\over 16\pi^2}\,{\mu\,A_t}{\tan\beta}\,
I(m_{\tilde{t}_1},m_{\tilde{t}_2},\mu)\,,
\label{eq:deltamblead}
\end{eqnarray}
where $I(m_1,m_2,m_3)$ (given in Ref.\cite{CGGJS})
is a slowly varying positive-definite function. We must 
emphasize that
the presence of such a leading term (and its expression) 
depends on the
renormalization scheme, that is, on the definition of 
$\tb$~(\ref{eq:deltabeta}).
In Fig.\,\ref{fig:deltatbmssm}a ($\mu<0$ scenario) the positive 
SUSY-QCD contribution compensates
the large negative QCD corrections, and thus the effect of the 
SUSY-EW sector
are clearly visible. There is a region (around $\tb\simeq50$) 
where the SUSY-QCD
correction fully cancels the QCD one, and we are left with only 
the SUSY-EW
correction. The $\mu>0$ scenario (Fig.\,\ref{fig:deltatbmssm}b) 
is very different. The
large negative SUSY-QCD corrections add up to the already 
large QCD ones, the
positive (due to the $\mu\,A_t<0$ constraint) EW corrections 
can not be large
enough to compensate for them.

\figmasthdm

Note that the Higgs contribution $\delta_{\rm Higgs}$ in 
Fig.\,\ref{fig:deltatbmssm}
is much smaller than the other ones. After imposing the SUSY 
couplings in the
vertex formulae of Sect.\,2 there is a large cancellation among 
the various
contributions. 
The reason can be seen in Fig.\,\ref{fig:deltamass2hdmI} where we
present the evolution of the corrections with $\tan\alpha$ and 
the pseudoscalar Higgs
boson mass. We see that the large corrections are attained for 
a specific
scenario: a definite value of $\tan\alpha$ 
(Figs.\,\ref{fig:deltamass2hdmI}a and c) and
large mass splitting (Figs.\,\ref{fig:deltamass2hdmI}b and d). 
These conditions
cannot be fulfilled in the MSSM, as $\tan\alpha$ and the 
mass splitting are
functions of the input parameters $\tb$ and $\mh$. The 
evolution of the
corrections with $\tan\alpha$ of the Type I $2$HDM
(Fig.\,\ref{fig:deltamass2hdmI}a) illustrates the general 
behaviour of the low
$\tb$ regime for both types of $2$HDM.

\figatmssm

In Fig.\,\ref{fig:deltaAtmssm} we present the evolution with the
Soft-SUSY-Breaking trilinear coupling of top squarks, which 
governs the behavior
of the SUSY-EW corrections~(\ref{eq:deltamblead}). We can 
see that they
effectively change sign with $A_t$, though the full 
correction deviates a bit of
the leading linear behaviour of eq.~(\ref{eq:deltamblead}). 
The shaded region
around $A_t\simeq0$ is excluded by the conditions on the 
squark masses.

\section{Implications for the Tevatron data}
Next we turn to the discussion of the dramatic implications 
that the EW
effects may have for the decay
$t\rightarrow H^+\,b$ at the Tevatron.  The original analysis 
of the data
(based on the non-observation of any excess of $\tau$-events) 
and its
interpretation in terms of limits on the $2$HDM parameter space
was performed in Ref.\cite{CDF} (for Type II models) without 
including
the EW corrections. In these references an
exclusion plot is presented in the
$(\tb,\mh)$-plane after correcting for QCD effects only.
The production cross-section of
the top quark in the  ($\tau$,$l$)-channel can be easily related to
the decay rate of $t\rightarrow H^+\,b$ and the branching ratio of
$H^+\rightarrow \tau^+\,\nu_{\tau}$ as follows:
\beq
\sigma_{l\tau}=\left[\frac{4}{81}\,\epsilon_1+\frac{4}{9}\,
{\Gamma (t\rightarrow H^+\,b)\over
\Gamma (t\rightarrow W^+\,b)}\,
BR(H^+\rightarrow \tau^+ \,\nu_\tau)\,\epsilon_2\right]\,
\sigma_{t\bar{t}}\,,
\label{eq:bfrac}
\eeq
with
\beq
BR(H^+\rightarrow \tau^+ \,\nu_\tau)=
\frac{\Gamma(H^+\rightarrow \tau^+\,\nu_\tau)}
{\Gamma(H^+\rightarrow \tau^+\,\nu_\tau)+\Gamma(H^+\rightarrow
c\,\bar{s})}\,\,, \eeq
where we use the QCD-corrected amplitude for the last term in the
denominator\cite{Gambino}.  
\figexcthdm

From Figs.\,\ref{fig:exc2hdm} and \ref{fig:excmssm} we 
inmediately see the
impact of the loop effects both in the general $2$HDM 
and the MSSM. We
have plotted the
perturbative exclusion regions in the
parameter space $(\tb,\mh)$ for intermediate and extreme 
sets of $2$HDM
inputs A, B, B' and C (Figs.\,\ref{fig:exc2hdm}a and b) 
and for the MSSM sets A
and B (Fig.\,\ref{fig:excmssm}). In Type I models 
(\ref{fig:exc2hdm}a) we see that the bounds obtained from
the EW-corrected amplitude are generally less restrictive than 
those obtained by
means of tree-level and  QCD-corrected amplitudes. Evolution 
of the excluded
region from set A to set C in Fig.\,\ref{fig:exc2hdm}a shows 
that the region tends to evanesce,
which is indeed the case when we further increase $M_{A^0}$ 
in set C.
In Type II models (\ref{fig:exc2hdm}b) we also show 
a series of possible scenarios.
We have checked that the maximum positive effect 
$\delta_{EW}>0$ (set A
in Fig.\ref{fig:exc2hdm}b) may completely cancel the QCD 
corrections and restore the full
one-loop width $\Gamma^{(II)}(t\rightarrow H^+\,b)$ to the 
tree-level
value just as if there were no QCD corrections
at all!
Intermediate possibilities (set B') are also shown. In the other 
extreme
the (negative) effects $\delta_{EW}<0$ enforce the exclusion 
region
to draw back to curve C where it starts to gradually
disappear into a non-perturbative corner of the parameter space 
where one
cannot claim any bound whatsoever!!.

\figexcmssm
In the MSSM we find a similar behaviour. For the first scenario 
($\mu<0$) the
large positive SUSY corrections take the exclusion region up to 
the tree-level
expectation. In the scenario characterized by $\mu>0$ the large 
negative
corrections take this region to too low values of $\mh$ and too 
high values of
$\tb$ and one cannot claim any bound on the $\tb-\mh$ plane.

\section{Conclusions}

In the MSSM case, the Higgs
sector is of Type II. However, due to supersymmetric restrictions 
in the
structure of the Higgs potential, there are large cancellations
between the one-particle-irreducible vertex functions, so that 
the overall
contribution from the MSSM Higgs sector to the  correction 
(\ref{eq:defdelta})
is negligible. In fact, we have checked that when we take 
the Higgs
boson masses
as they are correlated by the MSSM we obtain the same result.
Still, in the SUSY case there emerges a large effect
from the genuine sparticle sector, mainly from the SUSY-QCD 
contributions to
the bottom mass renormalization counterterm\cite{CGGJS},
which can be positive or negative because the correction flips sign 
with the
higgsino mixing parameter~(\ref{eq:deltamblead}). In contrast, for 
general (non-SUSY) Type II models the
bulk of the EW correction comes from large unbalanced contributions
from the
vertex functions, which can also
flip sign with $\tan\alpha$ (Cf. Fig.\,\ref{fig:deltamass2hdmI}c) 
-- a free parameter
in the non-supersymmetric case. Although the size and sign of the 
effects can
be similar for a general Type II and a SUSY $2$HDM, they should be
distinguishable since the large corrections are attained for very 
different
values of the Higgs boson masses\cite{CGSH}. 

We have demonstrated that in both cases (SUSY and general $2$HDM) 
the loop
effecs may completely distort the previous analyses presented by 
the Tevatron collaborations.

\section*{Acknowledgements}

The work of J.G. has been financed by a grant of the Comissionat
per a Universitats i Recerca, Generalitat de Catalunya (FI95-2125).  
This work has also been partially supported by CICYT under 
project No. AEN95-0882.


\end{document}